# Single Bessel tractor-beam tweezers

F. G. Mitri[a)]

*Chevron – Area 52 Technology, ETC, 5 Bisbee Court, Santa Fe, New Mexico 87508, USA*

**Abstract:** The tractor behavior of a zero-order Bessel acoustic beam acting on a fluid sphere, and emanating from a finite circular aperture (as opposed to waves of infinite extent) is demonstrated theoretically. Conditions for an attractive force acting in opposite direction of the radiating waves, determined by the choice of the beam's half-cone angle, the size of the radiator, and its distance from a fluid sphere, are established and discussed. Numerical predictions for the radiation force function, which is the radiation force per unit energy density and cross-sectional surface, are provided using a partial-wave expansion method stemming from the acoustic scattering. The results suggest a simple and reliable analysis for the design of Bessel beam acoustical tweezers and tractor beam devices.

## 1. Introduction

The ability to manipulate and trap particles in a confined space into desired patterns is a critical process in cell separation [1], tissue engineering [2] and biological scaffolding [3-5], the science of materials [6] and metamaterial composites [7], cell microarrays [8, 9], and other research areas.

Various modalities for trapping and manipulating particles using tweezers devices exist [10], including optoelectronic, magnetic, electrical, optical and acoustical tweezers (See the references list in Ref. [10]). However, due to some remaining limitations of these tools, significant interest is constantly directed toward the development of versatile, fast and cost-effective particle trapping devices.

Along this line of research, acoustical tweezers have been introduced as one of the most reliable and cost-effective devices, which use the forces of ultrasonic radiation [11] to trap and manipulate particulate matter. Substantial research [10, 12-15] has been focused on various applications since the earlier version used counter-propagating waves from two separate ultrasonic probes to form a beam of standing waves and trap microparticles [16].

Most established acoustical tweezers devices use counter-propagating waves [10, 12, 13, 17] to setup standing wave nodes and anti-nodes to trap the particles [7, 18-20]. Moreover, experimental [21, 22] and theoretical [23-26] research suggested the use of counter-propagating Bessel beams, to trap particles over

[a)]Electronic mail: F.G.Mitri@ieee.org

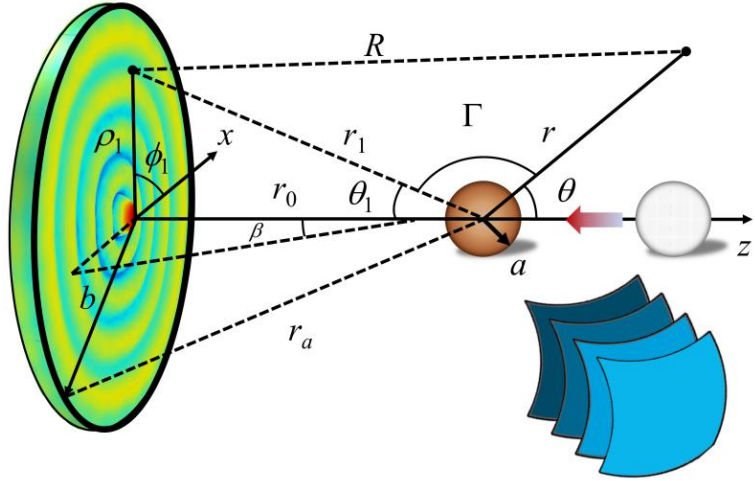

Fig. 1. A finite circular zero-order Bessel beam source with radius *b* emits acoustical waves, incident upon a spherical particle of radius *a* centered on its axis of wave propagation. The presence of the sphere causes the incident waves to scatter. The distance *r* denotes the position from the center of the sphere to an observation point. *R* denotes the distance from a point on the piston's surface to the observation point, and $r_0$ is the distance from the center of the radiator to the center of the sphere.

extended distances. Such methods, however, may require complex equipment for transducer design, and may be time-consuming for calibration purposes. In addition, the counter-propagating (dual-wave) setup often creates multiple trapping points [7, 27], which could be problematic when only a single particular region for particle trapping and manipulation is needed in the media.

Motivated by the need to further develop faster, simpler and more reliable tools in acoustical tweezers, a *single*-beam of focused progressive waves had effectively produced lateral trapping in the off-axial direction [28]. Moreover, a single-beam particle trapping device with an extended focal region [29] achieved trapping by emulating the behavior of an axicon using a multi-foci air-backed Fresnel lens, while a recent theoretical approach proposes the use of a single piezo-disk radiator with uniform vibration (which is readily available commercially) to accomplish particle entrapment in the near-field zone of the piston source [30].

In contrast to counter-propagating beams, single-Bessel acoustical beams [31, 32] have been suggested for particle trapping and tweezing [33-35], which required fabricating a 16-element circular array device by dicing a piezo-ceramic ring, backed with an absorbing layer of mixed epoxy and alumina [34]. Other methods, such as the non-uniform poling technique [31, 36], or a Bessel-driven multi-annular transducer [37], have been suggested and proven to produce a finite (limited-diffracting) Bessel beam. Nevertheless, applying such approaches in the design and manufacturing of Bessel acoustical tweezers can be laborious, costly, and may require advanced equipment and extensive hardware development. Therefore, it is of some importance to work out improved and reliable methods for the experimental verification and theoretical predictions of acoustical tweezing capabilities using conventional (commercially-available) ultrasonic probes operating with Bessel modes.

This analysis suggests the use of a standard piezo-disk transducer vibrating *non-uniformly* according to one of its radially symmetric modes [38-43] to generate a zero-order Bessel beam [44] (Fig. 1), having a pressure maximum in amplitude (or intensity) along its center (as opposed to the higher order Bessel beams having an axial null [45]). For these radial modes, there exist vibrations (and subsequent acoustic radiation) in both radial and axial (out-of-plane) directions, which can be in or out of phase depending on the mode order (See Ch. 5 in [46], and the animations in [44]). In the radial direction, there are nodal circles over the thickness of the disk at well-defined radii at which the displacement in the radial direction is zero, and the number of nodal circles increases as the mode order increases [46]. Those can be closely approximated by a normal velocity profile at the surface of the vibrating radiator ($z = 0$) in the form of a cylindrical Bessel function of order zero (denoted by $J_0$), having a maximum in amplitude at the center of the beam.

In contrast with the studies using Bessel waves of infinite extent [Eq.(19) in Ref. [23], [47]] which carry an infinite amount of energy and thus, are practically and physically unrealizable, the aim here is to investigate the prediction of a pulling (negative) force for particle tweezing using a *single*-Bessel beam generated from a *finite* circular radiator driven at one of its radially-symmetric modes. For particle trapping in real-world applications, this approach may be the most simpler, effective and rapid method for the generation of an ultrasonic Bessel beam of limited-diffraction by exciting the piezoelectric crystal (or ceramic) disk at one of its radially-symmetric vibrational modes, so as to achieve a single Bessel tractor beam tweezers device.

Through numerical simulations, the acoustic radiation force function for a fluid hexane sphere, which is the radiation force per unit energy density and cross-sectional area, is evaluated. The fluid sphere example is of particular interest in various bioengineering/biophysical/(bio)chemical applications, which may closely mimic the behavior of a single cell or a liquid droplet in a Bessel beam. Nevertheless, the analysis can be readily extended to elastic or viscoelastic (layered) spheres [48, 49], or shells [50, 51], providing their appropriate scattering coefficients are used. In this analysis, the sphere is immersed in an ideal (non-viscous) fluid, and particular emphasis is given on the distance separating the sphere from the Bessel acoustic source $r_0$, the radius of the transducer $b$, and the half-cone angle of the beam $\beta$. The numerical computations are of particular importance in the rapid, efficient and practical design of single-Bessel beam tweezers since they allow exploring a broad range of parameters and anticipating conditions for which particle trapping is expected to occur.

## 2. Theoretical analysis

Consider a circular piezo-disk transducer of radius $b$ centered on a spherical fluid particle radius $a$ immersed in a non-viscous fluid. The distance from the center of the sphere to the center of the radiator is $r_0$. Assuming a *non-uniform* vibration velocity at the surface of the circular radiator in the form of $v|_{z=0} = V_0 J_0(k_\rho \rho_1)$, (where $k_\rho = k \sin\beta$ is the radial wave-number and $k$ is the wave-number of the acoustic radiation, $\beta$ is the half-cone angle of the beam, $\rho_1$ is the distance from the center of the radiator to a point in the transverse plane, and $J_0(.)$ is the cylindrical Bessel function) the incident acoustic velocity potential field is obtained from the Rayleigh integral as [52]

$$\Phi_i = \frac{1}{2\pi} \iint_{S_r} \frac{V_0 J_0(k_\rho \rho_1) e^{i(kR-\omega t)}}{R} dS_r, \qquad (1)$$

where $R$ is the distance from the observation point to the finite source of circular surface $S_r$ (Fig. 1).

Using the addition theorem for the spherical functions, i.e. (10.1.45) and (10.1.46) in Ref. [53] such that $r \leq r_1$, Eq.(1) is expressed as

$$\Phi_i = \frac{ikV_0 e^{-i\omega t}}{2\pi} \sum_{n=0}^{\infty} (2n+1) j_n(kr) \iint_{S_r} h_n^{(1)}(kr_1) J_0(k_\rho \rho_1) P_n(\cos\Gamma) dS_r, \qquad (2)$$

where $j_n(\cdot)$ and $h_n^{(1)}(\cdot)$ are the spherical Bessel and Hankel functions of the first kind, $P_n(\cdot)$ are the Legendre functions, and the differential surface $dS_r = \rho_1 d\rho_1 d\phi_1 = r_1 dr_1 d\phi_1$, since $r_1^2 = \rho_1^2 + r_0^2$ (Fig. 1).

Using the addition theorem for the Legendre functions[54, 55] and the definition of the angles as given in Fig. 1, the Legendre functions $P_n(\cos\Gamma)$ can be expressed as

$$P_n(\cos\Gamma) = \sum_{\ell=0}^{n} (2-\delta_{\ell,0}) \frac{(n-\ell)!}{(n+\ell)!} (-1)^{n+\ell} P_n^\ell(\cos\theta) P_n^\ell(\cos\theta_1) \cos\ell(\phi-\phi_1), \qquad (3)$$

where $\delta_{ij}$ is the Kronecker delta function, and $P_n^\ell(\cdot)$ are the associated Legendre functions. After the integration Eq.(3) with respect to $\phi_1$, and substituting the result into Eq.(2), the incident velocity potential can be expressed as [44]

$$\Phi_i = \Phi_0 \sum_{n=0}^{\infty} \Lambda_{J_0,n} i^n (2n+1) j_n(kr) P_n(\cos\theta) e^{-i\omega t}, \qquad (4)$$

where,

$$\Lambda_{J_0,n} = i^{n+1} \int_{kr_0}^{kr_a} (kr_1) h_n^{(1)}(kr_1) J_0\left(k_\rho \sqrt{r_1^2 - r_0^2}\right) P_n\left(\frac{r_0}{r_1}\right) d(kr_1), \qquad (5)$$

and $\Phi_0 = V_0/k$.

For the standard case of an acoustical field emanating from a circular radiator freely vibrating with *uniform* velocity [56, 57] (like a rigid piston) or simply supported[58], a simplified form of the integral given in Eq.(5) has an exact closed-form solution [30] that was initially provided assuming a time dependence in the form of $e^{i\omega t}$ [56]. This case can also be recovered by setting $\beta = 0°$ (or $k_\rho = 0$) in Eq.(5). Moreover, when the half-cone angle $\beta$ is smaller than 1 radian (=180°/$\pi$ ~ 57.3°) and the partial-wave number $n \gg 1$, the simplest case of the Mehler–Heine formula ($\mu = 0$ in Eq.(9.1.71) in [53]) can be used so that the cylindrical Bessel function in Eq.(5) may be approximated as

$$J_0\left(k_\rho \sqrt{r_1^2 - r_0^2}\right) \approx P_n(\cos\beta), \qquad (6)$$

provided that [59]

$$\left(r_1^2 - r_0^2\right)^{\frac{1}{2}} \approx \left(n + \tfrac{1}{2}\right)/k. \qquad (7)$$

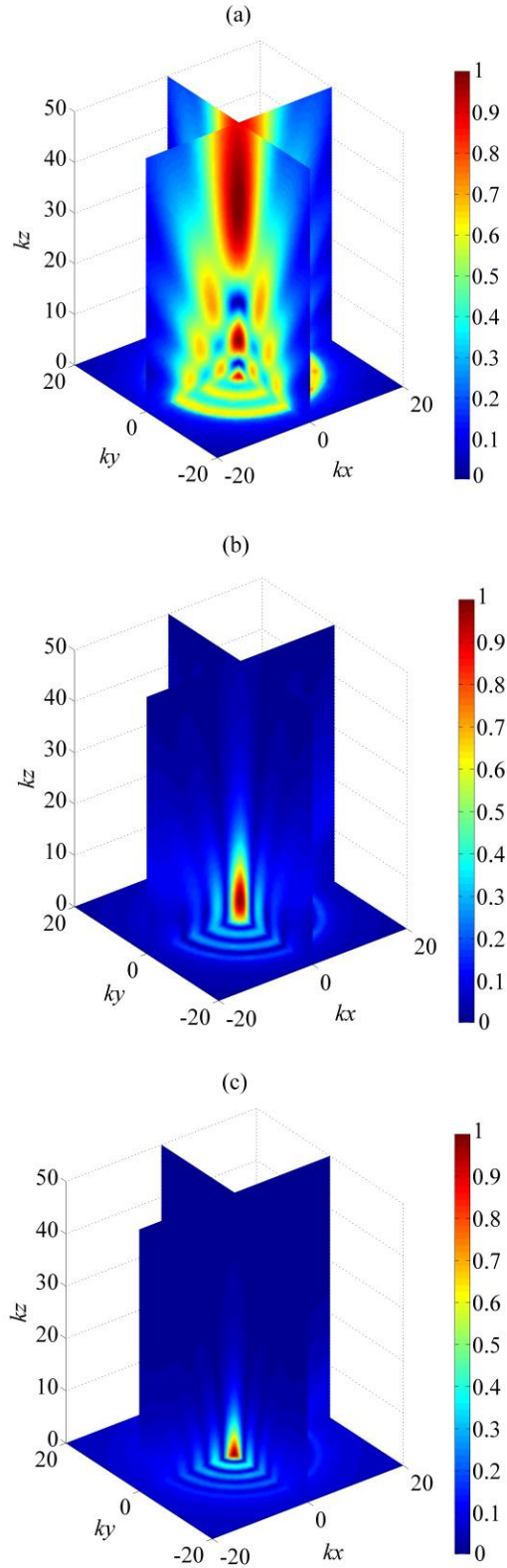

Fig. 2. Normalized magnitudes of the incident velocity potential field representing the radiated field from a finite transducer for $kb = 15$ with uniform vibration $\beta = 0°$ (a), and Bessel beam profiles for $\beta = 45°$ (b), and $\beta = 65°$ (c), where $b$ is the transducer's radius. The numerical computations were performed in the transverse dimensionless ranges $-20 \leq (kx, ky) \leq 20$, and the axial dimensionless range $0 < kz \leq 50$. The transverse magnitude plots were evaluated at $kz = 1$.

As noticed by inserting Eq.(6) into Eq.(5), a closed-form solution for $\Lambda_{J_0,n}$ is obtained as

$$\Lambda_{J_0,n}\Big|_{\substack{n\to\infty \\ \beta\to 0}} = i^{n+1} P_n(\cos\beta) f_n, \qquad (8)$$

where the closed-form expression for $f_n$ is explicitly given by Eq.(9) in Ref. [30]. Note also that for the case of waves of infinite extent, the function $f_n = i^{-n} e^{ikr_0}$, [30, 60], thus,

$$\Lambda_{J_0,n}^{\infty}\Big|_{\substack{n\to\infty \\ \beta\to 0}} = i P_n(\cos\beta) e^{ikr_0}. \qquad (9)$$

Substituting Eq.(9) into Eq.(4) leads to the known partial-wave expansion for the incident velocity potential field of a zero-order Bessel beam of progressive waves[23].

In addition, note that Eq. (7) is the standard localization principle relationship [61] in the case of plane wave scattering for which each partial-wave $n$ is associated with a ray (or a bundle of rays) located at a distance $(r_1^2 - r_0^2)^{\frac{1}{2}}$ of the propagation axis passing through the center of the scatterer. Thus, using Eq.(8), the series in Eq.(4) characterizing the incident waves may be interpreted in terms of geometrical acoustical rays.

For the purpose of the present investigation, Eq.(5) is evaluated numerically as no closed-form solution is currently available.

The presence of the sphere in the beam's path produces a scattered wave, for which the axial scattered velocity potential is expressed as

$$\Phi_s = \Phi_0 \sum_{n=0}^{\infty} \Lambda_{J_0,n} i^n (2n+1) S_n h_n^{(1)}(kr) P_n(\cos\theta) e^{-i\omega t}, \qquad (10)$$

where $S_n$ are the scattering coefficients for a fluid sphere, which are explicitly given in Ref. [62]. Thus, the total (incident + scattered) velocity potential field, which can be used to evaluate the acoustic radiation force, is the summation of Eqs.(4) and (9).

The acoustic radiation force can be evaluated through the dimensionless function $Y$, which is the radiation force per unit energy density and unit cross-sectional surface [63]. In the case of a finite Bessel beam, the radiation force function is expressed as

$$Y = Y_{J_0}$$
$$= -\frac{4}{(ka)^2} \sum_{n=0}^{\infty} (n+1) \begin{bmatrix} \mathrm{Re}\left(\Lambda_{J_0,n}\Lambda_{J_0,n+1}^*\right)(\alpha_n + \alpha_{n+1} + 2\alpha_n\alpha_{n+1} + 2\beta_n\beta_{n+1}) \\ + \mathrm{Im}\left(\Lambda_{J_0,n}\Lambda_{J_0,n+1}^*\right)(\beta_{n+1} - \beta_n + 2\alpha_n\beta_{n+1} - 2\beta_n\alpha_{n+1}) \end{bmatrix}, \qquad (11)$$

where $\alpha_n = \mathrm{Re}(S_n)$, $\beta_n = \mathrm{Im}(S_n)$, and the superscript * denotes a complex conjugate.

In the infinite Bessel beam case, the substitution of Eq.(9) into Eq.(11) leads to the well-known expression for the radiation force function (Eq.(19) in [23])

$$Y = Y_{J_0}^{\infty}$$
$$= -\frac{4}{(ka)^2} \sum_{n=0}^{\infty} (n+1)[\alpha_n + \alpha_{n+1} + 2\alpha_n \alpha_{n+1} + 2\beta_n \beta_{n+1}] P_n(\cos\beta) P_{n+1}(\cos\beta). \quad (12)$$

## 3. Numerical results and discussion

The accuracy of the numerical integration method and its validation with a closed-from solution for a known beam is performed first. In the case of a finite circular piston radiator vibrating uniformly, the parameter $\Lambda_{J_0,n} = \Lambda_n$ has an exact closed-form solution[30], and used here to advantage. The exact solution, given explicitly by Eq.(9) in Ref. [30], is computed and compared with the results of the numerical integration of Eq.(5) for $\beta = 0°$, corresponding to the case of plane waves, with a chosen sampling step of $\delta(kr_1) = 10^{-4}$. With this chosen value of $\delta(kr_1)$, the absolute error for a range of chosen sets $(kr_0, kr_b)$ is below $10^{-7}$. Thus, a smaller sampling step is not required since the absolute error is insignificant.

First, sample computations of the radiated field from a finite Bessel transducer with $kb = 15$ in lossless water were performed by numerical evaluation of the steady-state (i.e. time-independent) incident velocity potential field as given by Eq.(4). The series in Eq.(4) were truncated following the convergence criterion; $|\Phi_{n,i} - \Phi_{n-1,i}|/|\Phi_{n-1,i}| \ll \varepsilon \sim 10^{-6}$, where $\Phi_{n,i}$ is the incident velocity potential estimate obtained by truncation of the exact series expansion after the $n$th term. The value of $kr_0$ was selected so that the maximum value of $|kr_0 - kz| < 70$ [56]. Three cases were considered; the first corresponding to a finite radiator with uniform vibration (i.e. $\beta = 0°$). For the uniform vibration case, the results were compared with those obtained previously[56]. Perfect agreement was observed at every field point tested. The second and third cases correspond to finite Bessel beams with $\beta = 45°$ and $\beta = 65°$ chosen as illustrative examples. The related results are displayed in Fig. 2-(a)-(c), corresponding to $\beta = 0°$, 45° and 65°, respectively. From Fig. 2, deviations from the uniform-vibration case are clearly noticeable for $kb = 15$; the successive maxima and minima in the magnitude of the incident velocity potential (or pressure) field that occur in the near-field in the absence of the scatterer for $\beta = 0°$, are not manifested for Bessel beams. Moreover, the region over which the beam is of limited-diffraction contracts, and the maximum in magnitude moves closer to the radiator's surface when $\beta$ increases.

Now, the case of a fluid hexane spherical drop ($\rho_{hex.} = 656$ kg/m$^3$, $c_{hex.} = 1078.5$ m/s) immersed in non-viscous water ($\rho_0 = 1000$ kg/m$^3$, $c_0 = 1500$ m/s) is considered. The radiation force function [as given by Eq.(11)] is evaluated for three distinct values of the dimensionless Bessel source radius $kb$ (= 15, 30 and 45, respectively), ranging from a small to a relatively large dimensionless radius. The computations are performed in the range $0° \leq \beta \leq 90°$ for $0 < kr_0 \leq 200$. An essential condition is also applied for the radiation force function computations; that is, the distance $r_0$ should always be larger than the sphere's radius $a$, which explains the "tilt" occurring in the plots of Fig. 3 as $ka$ increases.

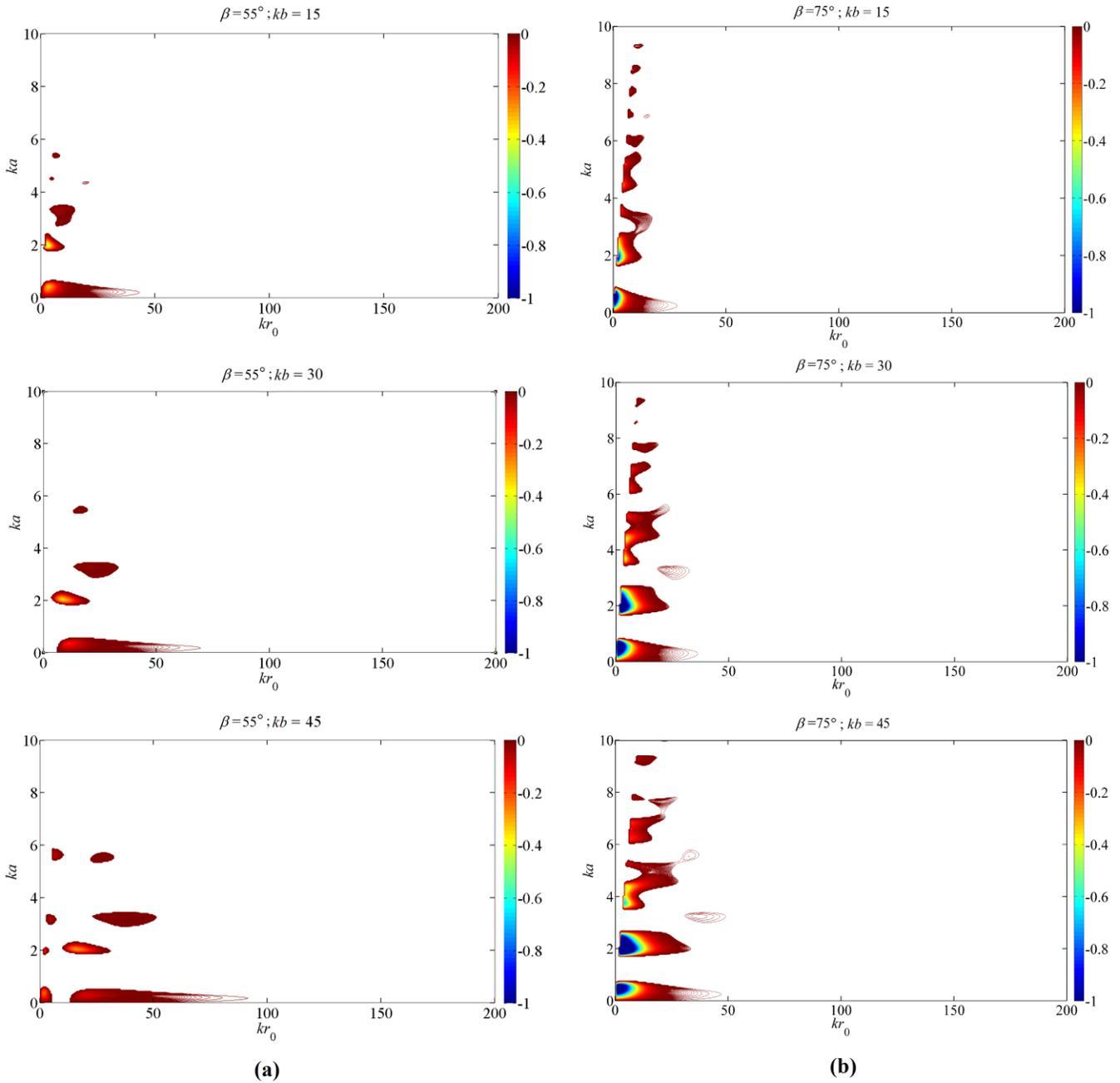

Fig. 3. Regions, where $Y_{J_0}$ is computed by Eq.(11) to be negative, are shown as well-defined "islands" for $\beta = 55°$ [panels in column (a)], and $\beta = 75°$ [panels in column (b)] for $kb$ = 15, 30 and 45, respectively. See also the Supplementary Animations 1-3 (http://dx.doi.org/10.1016/j.wavemoti.2014.03.010), corresponding to the parameter $kb$ = 15, 30 and 45, respectively. The color axis scaling is set to a minimum value of -1, and a maximum value of 0 for all panels for enhanced visualization.

The panels in Fig. 3-(a),(b) show the regions or "islands" where the radiation force function $Y_{J_0}$ is computed (for the hexane sphere) by Eq.(11) to be negative for $\beta = 55°$ [Fig. 3-(a)], and for $\beta = 75°$ [Fig. 3-(b)], for $kb = 15$ [first row], 30 [second row], and 45 [third row]. Moreover, the effect of changing $\beta$ is displayed in Supplementary Animation 1 for $kb = 15$, which shows that the regions over which $Y_{J_0} < 0$ for $\beta < 38°$, are limited to low $ka$ ($< 1$), whereas beyond this angle (i.e. $\beta \geq 38°$), a number of "islands" emerges, close to the Bessel beam radiator, for $kr_0 < 50$. Note also that the maximum negative force on a hexane fluid sphere arises for $ka \leq 2$, with the Bessel source being very close to the particle.

The effect of increasing the dimensionless radius on the negative force is displayed in the second row of Fig. 3-(a),(b). The comparison of the plots in the panels of the first row with those displayed in the second row (for the same half-cone angle value) shows that the regions over which $Y_{J_0}$ is negative are larger for all $\beta$'s. This is further demonstrated in the Supplementary Animation 2, which shows the effect of changing $\beta$ on the negative force acting upon a hexane sphere.

An additional increase of the dimensionless radius and its effect on the negative force is shown in the third row of Fig. 3-(a),(b). The Supplementary Animation 3 shows the effect of varying $\beta$ and the appearance of larger "islands" which are generally extended over larger $kr_0$ values. After comparing the results in the Supplementary Animations 1-3, one notices that the number of islands increases as $kb$ increases. This is further demonstrated in the Supplementary Animation 4, which shows the variations of the radiation force function versus $ka$ and $kr_0$ in the range $15 \leq kb \leq 45$ for $\beta = 75°$. Those results support the conclusion that increasing the transducer's radius increases the regions over which the radiation force is negative. Moreover, when $\beta$ increases, the negative force becomes strongly manifested closer to the Bessel acoustic radiator surface as the "islands" shift to lower $kr_0$ values.

The main result given in Eq.(5), used in Eq.(11) to compute the radiation force function for an isotropic fluid sphere centered on a Bessel beam radiator emerging from a finite circular transducer, allows treating the case where the absorption of the acoustic energy inside the droplet cannot be neglected. This is often the case for oil particles [64], or polymer-type [65] spheres. As a general statement, absorption by the sphere is shown to *degrade* the negative force in the context of Bessel beams [See Fig. 8 in Ref. [23]] and for a polyethylene viscoelastic sphere centered on a piston radiator vibrating uniformly[30]. In the numerical examples presented here, viscosity inside the hexane sphere has been neglected (as is the case for low viscosity hydrocarbon liquid drops), and the present results remain valid for fluid droplets with low viscosity.

Regarding the case where the spherical droplet may be placed off-axially with respect to the incident beam's axis, recent works [66-69] have extended the scattering theory formalism [70] and considered the general case of arbitrary incidence. The methods for the calculation of the radiation force [71, 72] and torque [64], which are based on the generalized scattering formalism [66-69], are applicable to *any beam* composed of acoustic (scalar) waves, and the present analysis can assist in extending the result to evaluate the off-axial (transverse) force and torque components on a sphere in a finite Bessel beam.

An additional comment should be made here regarding the effects of streaming forces and thermo-viscous effects. In applications in acoustofluidics and particle manipulation in small channels, previous investigations [73-75] predicted negative forces on a particle immersed in a viscous fluids [76], or when the particle (immersed in a non-viscous fluid) is close to a boundary [77]. It is important to note that for fluids with low viscosities, experiments have given satisfactory agreement with the non-viscous radiation force predictions using progressive waves incident upon an elastic sphere [78, 79] and an elastic cylinder [80]. In general, when the sphere is small compared to the acoustic wavelength (i.e. the viscous boundary layer thickness is comparable to the sphere's radius or larger), the viscous corrections are required. Otherwise, the dissipative effects are negligible. In the present analysis, the (Stokes) viscous boundary layer is assumed to be much smaller than the sphere's radius so the thermoviscous corrections are considered negligible.

## 4. Summary

In conclusion, this work demonstrated theoretically the existence of a negative pulling radiation force on a fluid hexane sphere of arbitrary size using a single-beam Bessel ultrasonic radiator as opposed to waves of infinite extent. Radiation form function predictions showed that the acoustical waves act as a "tractor" beam by inducing a pulling force on the particle, depending on the half-cone angle and distance from the Bessel source. Increasing the Bessel transducer radius increases the number of regions over which the acoustical waves act as a tractor beam. Should the results of an experimental setting prove to be in favor of the theoretical findings, this research could be very promising in the optimal design of Bessel acoustical tweezers devices in acoustofluidics and particle trapping.

**Acknowledgement**

The author acknowledges Dr. C. Pantea from the Los Alamos National Laboratory (LANL, MPA-MSID) for helpful discussions related to the radially-symmetric vibrational modes of ultrasonic transducers and for providing thorough references on this topic.